# Evidence for Three-Dimensionality in the Fermi Surface Topology of Layered Electron Doped Ba(Fe$_{1-x}$Co$_x$)$_2$As$_2$ Iron Superconductors

P. Vilmercati[1,2], A. Fedorov[3], I. Vobornik[4], Manju U.[4], G. Panaccione[4], A. Goldoni[5], A. S. Sefat[6], M. A. McGuire[6], B. C. Sales[6], R. Jin[1,6], D. Mandrus[6], D. J. Singh[6] and N. Mannella[1,*]

[1]Department of Physics and Astronomy, The University of Tennessee, Knoxville, TN 37996, USA
[2]Dipartimento di Fisica, Università degli Studi di Trieste, I- 34127, Trieste, Italy.
[3]Advanced Light Source, Lawrence Berkeley National Laboratory, Berkeley, California 94720, USA
[4]TASC National Laboratory, INFM-CNR, SS 14, km 163.5, I-34012 Trieste, Italy.
[5]Sincrotrone Trieste S.C.p.A., Area Science Park, S.S. 14, km 163.5, I-34012 Trieste, Italy.
[6]Materials Science and Technology Division, Oak Ridge National Laboratory, Oak Ridge, TN 37831, USA.
* Corresponding Author. E-mail: nmannell@utk.edu

## ABSTRACT

The electronic structure of electron doped iron-arsenide superconductors Ba(Fe$_{1-x}$Co$_x$)$_2$As$_2$ has been measured with Angle Resolved Photoemission Spectroscopy. The data reveal a marked photon energy dependence of points in momentum space where the bands cross the Fermi energy, a distinctive and direct signature of three-dimensionality in the Fermi surface topology. By providing a unique example of high temperature superconductivity hosted in layered compounds with three-dimensional electronic structure, these findings suggest that the iron-arsenides are unique materials, quite different from the cuprates high temperature superconductors.

The recent discovery of high-temperature superconductivity in Fe-based superconductors (FeSC) has provided the opportunity of studying high-temperature superconductivity and its relation to magnetism in a wide range of magnetic element-based materials [1]. Cuprate high-temperature superconductors (HTSC) and FeSC materials are the only two families of compounds with superconducting critical temperature ($T_C$) known to exceed 55 K. It is therefore important to determine whether and to what extent these two classes of materials are similar. Much like in cuprates, superconductivity in FeSC emerges in close proximity to a long-range-ordered antiferromagnetic ground state, indicating that the interplay between superconductivity and spin-density-wave instability is likely important in FeSC [2,3,4,5]. On the other end, spectroscopic evidence has shown that the electronic structure of both the superconducting and normal state is quite different from that of cuprates [6,7,8]. Despite their layered structure, (Ba,K)Fe$_2$As$_2$ materials have recently been reported to exhibit superconducting properties that are in fact quite isotropic, a behavior drastically different compared to that of other layered superconductors [9]. It has been proposed that the nearly isotropic critical field of (Ba,K)Fe$_2$As$_2$ might be linked to its distinctive three-dimensional electronic structure and Fermi surface (FS) topology, indicating that the reduced dimensionality is not a prerequisite for high temperature superconductivity, contrary to what has been suggested for cuprate HTSC. A direct determination of the dimensionality of the electronic structure and FS topology in FeSC materials is thus timely and important, as it is expected to provide firm grounds of comparison for establishing commonalities and differences with the cuprate HTSC.

In this Communication, we address this issue by reporting the results of Angle Resolved Photoemission Spectroscopy (ARPES) measurements of the electronic structure in the normal state of BaFe$_2$As$_2$ and BaFe$_{1.8}$Co$_{0.2}$As$_2$ single crystals. The data reveal a marked three-dimensional character of the electronic band structure and FS topology for both the parent and doped compounds.

The Ba(Fe$_{1-x}$Co$_x$)$_2$As$_2$ materials crystallize in high quality single crystals with the tetragonal ThCr$_2$Si$_2$ structure-type (space group I4/mmm), and have a maximum T$_C$ = 22K for x = 0.1 nominal doping [10]. Electron probe microanalysis carried out with a JEOL JSM-840 scanning electron microprobe on several crystals indicated that 8.0(5)% of the Fe is replaced by Co in BaFe$_2$As$_2$. This composition will be represented as the nominal BaFe$_{1.8}$Co$_{0.2}$As$_2$ or 8 % Co doping. Co-doping of the BaFe$_2$As$_2$ system is advantageous since electronic carriers are added directly in the FeAs planes, and Co is easier to handle than alkali metals. In remarkable contrast to cuprates HTSC, the BaFe$_2$As$_2$ system appears to tolerate considerable disorder in the FeAs planes. High-quality single crystals were grown out of FeAs flux according to modalities described elsewhere [10]. The ARPES measurements were carried out on Bl. 10.0.1 and Bl. 12.0.1 at the Advance Light Source, and the low-energy branch of the APE-INFM beamline in Elettra Synchrotron (Trieste, Italy). Several samples from different batches have been measured at 30 K after being cleaved in situ in a pressure better than 3×10$^{-11}$ Torr. The total instrumental energy resolution was 12-25 meV, while the angular resolution was set to either ± 0.5° or ± 0.1°, which correspond to ≈ 0.05 Å$^{-1}$ and ≈ 0.01 Å$^{-1}$, respectively.

The basic features of the electronic structure of the BaFe$_2$As$_2$ materials have been exposed by several ARPES investigations [6,11,12,13,14,15,16,17]. The electronic structure consists of disconnected FS, a hole-like pocket at the zone center (Γ), and an electron-like pocket at the zone corner (M), with the electronic bands exhibiting some renormalization effects. A direct comparison of spectra collected along the ΓX direction (i.e. [100]) in the parent BaFe$_2$As$_2$ and doped BaFe$_{1.8}$Co$_{0.2}$As$_2$ ( T$_C$ = 22 K ) compounds is shown in Fig. 1. Since these spectra have been collected in identical experimental conditions (same photon energy, in-plane photon polarization), the differences in the spectral features reflect intrinsic differences between the parent and doped compound caused by doping. Upon Co substitution, the hole pocket at Γ becomes smaller, while the electron pocket at M becomes larger (Fig. 1). These changes are indicative of an upward shift of the chemical potential, as expected for electron doping. Similar observations occur both when using different photon energies and when spectra are collected along the ΓM direction, thus indicating that Co substitution is indeed effective in doping the FeAs plane with electrons. Our result confirms the predictions of density functional calculations, namely that the Co doped materials behave like a coherent alloy with the main effect induced by Co doping being an upward shift of the chemical potential [10]. This "rigid-band" behavior contrasts what would be expected in a strongly correlated system, thus providing another distinction between FeSC and cuprates.

ARPES is a premiere tool for measuring energy dispersion and symmetry of bulk bands [18]. For electron bands dispersing along the direction perpendicular to the surface plane (the c axis), the allowed direct transitions will shift in energy and consequently in momentum perpendicular to the sample plane ($k_z$) when the photon energy is changed [18]. Signatures of three-dimensional dispersion are thus typically revealed in ARPES experiments by changes of the valence band peaks position when the spectra are collected in normal emission while changing the photon energy, as shown in Fig. 2a. Besides the enhancement of the structure close to the Fermi level ($E_F$) resulting from the Fe 3p→Fe3d resonance at hν ≈ 56 eV, the spectra show dispersion of a structure with binding energy (BE) of ≈ 600 meV, a distinct signature of the three dimensional character of this band. Also visible is another structure at 200 meV BE with weaker dispersion. The modulation of these structures exhibits a periodicity as a function of photon energy which, when translated into the $k_z$ quantum number using a free electron final state approximation with an inner potential of 15 eV [18], corresponds to ≈ 1 Å$^{-1}$, a value remarkably close to the expected periodicity along $k_z$ ≈ 4π/c = 0.968 Å$^{-1}$, with c = 12.98 Å [10].

To address the possible three-dimensional character of the FS, we have carried out a systematic photon energy dependence of the states close to $E_F$ both at the Γ point and at the M point. The data have been analyzed with Momentum Distribution Curves (MDC), which identify as Fermi wavevectors the maximum positions of the momentum distribution of spectral weight at $E_F$. The MDC analysis of the spectra collected in BaFe$_{1.8}$Co$_{0.2}$As$_2$ at the M point (Fig. 2b) show little changes of the Fermi crossing point, i.e. zero binding energy, for different photon energies, indicating that the electron pockets are cylinder-like with a low degree of warp, as predicted by LDA calculations [10].

Fig. 3 shows spectra collected in BaFe$_{1.8}$Co$_{0.2}$As$_2$ at the Γ point along the ΓM direction [100] for different photon energies and high momentum resolution (≈ 0.01 Å$^{-1}$). The most noteworthy characteristic of these data is the marked dependence of the Fermi crossing points on the photon energy, which reveals a strong dispersion along the $k_z$ direction in momentum space. Particularly interesting is also the observation of a band located ≈ 200 meV below $E_F$. This band is predicted by LDA calculations when atom positions are obtained by relaxing the internal coordinates of the As vertical position using LDA total energy minimization, but not when atom positions are taken from structural refinement data [19]. Changes in As atom positions of the order of 0.1 Å may occur as a result of a surface relaxation, a possibility which warrants further investigations.

In the case of well known two dimensional systems like the layered manganites, layered cuprates and organic superconductors, the FS are very two-dimensional, with a cross-sectional area that varies little along the interlayer direction or, analogously, with Fermi crossing points which are insensitive to the use of different photon energies [20,21]. This occurrence does not seem to be the case in BaFe$_{1.8}$Co$_{0.2}$As$_2$. To better illustrate this point, Fig. 4a shows the MDC analysis of the spectra shown in Fig. 3. The MDC curves make it apparent how the Fermi crossing points change markedly along the ΓX direction when the photon energy is changed. These data show that there is dispersion along the $k_z$ direction, indicating a three-dimensional character of the FS topology in the FeSC compounds. LDA calculations have predicted that in Ba(Fe$_{1-x}$Co$_x$)$_2$As$_2$ compounds the FS around the Γ point consists of a cylinder with small cross sectional area that flares out in proximity of the Z point [10]. Fig. 4b shows the distribution of spectral weight at $E_F$ in color scale as a function of the in-plane momentum along ΓX ($k_x$) with the out-of-plane momentum $k_z$ displayed on the vertical axis. This plot represents a cross sectional view of the FS profile, and illustrates how considerable the dispersion along the $k_z$ direction in momentum space is, manifested as a strong variation of the cross-sectional cut along $k_x$ as $k_z$ is changed, fully consistent with the LDA predictions as shown in the same figure. The three-dimensional character of the FS topology in BaFe$_{1.8}$Co$_{0.2}$As$_2$ provides a rationale for the unusual nearly isotropic critical field of (Ba,K)Fe$_2$As$_2$ [9]. Our results are also consistent with ARPES data collected with photon energies of 50 eV and 80 eV in ref. 17, in which similar Fermi crossing points were found at these two energies. This is in agreement with our data and analysis, since photon energies of 50 and 80 eV correspond to similar values of $k_z$ [22].

We have carried out similar experiments using different light polarizations. We observed that with in-plane photon polarization it was impossible to detect the band at 200 meV, most likely a result due to matrix elements effects, which have been shown to be quite strong in the FeSC materials [12]. More importantly, the strong $k_z$ dispersion effects shown in Fig. 3 remain observable, independent of the polarization of the incoming light. We also found that the three dimensionality of the FS topology is also a characteristic of the parent compound. Fig. 5 show data collected in $BaFe_2As_2$ at the $\Gamma$ point, along the $\Gamma M$ direction, at different photon energies, with in-plane photon polarization. The marked dependence on the photon energy of the Fermi crossing points is substantial, as much as that of the doped compound shown in Fig. 3.

The FeSC show an intimate connection between magnetism and superconductivity. Experiments show that the undoped, spin density wave ground state of the parent compound is metallic, as the normal states of the superconducting phases. This is in contrast to cuprates, whose undoped ground state is a Mott insulator. In cuprates, the connections between the undoped Mott insulators and the superconducting phases occur only at high energy, i.e. in the Hubbard bands. Our results further underscore the connection in FeSC between the magnetic and superconducting phases by showing that they have a similar degree of three-dimensionality.

In conclusion, it appears that FeSC are quite unique high temperature superconductors, with an itinerant Fe d-band character, high density of states at $E_F$, and capable of hosting high temperature superconductivity without the signatures of strong local Mott-Hubbard type correlations that characterize cuprate HTSC. Furthermore, the present results show by the example of $Ba(Fe_{1-x}Co_x)_2As_2$ compounds that the FeSC materials have a three-dimensional FS topology, thus indicating that reduced dimensionality is not a necessary condition for high temperature superconductivity. These results assist in establishing a unique character of the FeSC materials, which should be considered a new class of high temperature superconductors, quite unlike the cuprates.

The work at the ALS and Elettra is supported by NSF grant DMR-0804902. The work at Oak Ridge is sponsored by the Division of Materials Science and Engineering, Office of Basic Energy Sciences. Oak Ridge National Laboratory is managed by UT-Battelle, LLC, for the U.S. Department of Energy under Contract No. DE-AC05-00OR22725. Portions of this research performed by Eugene P. Wigner Fellows at ORNL.

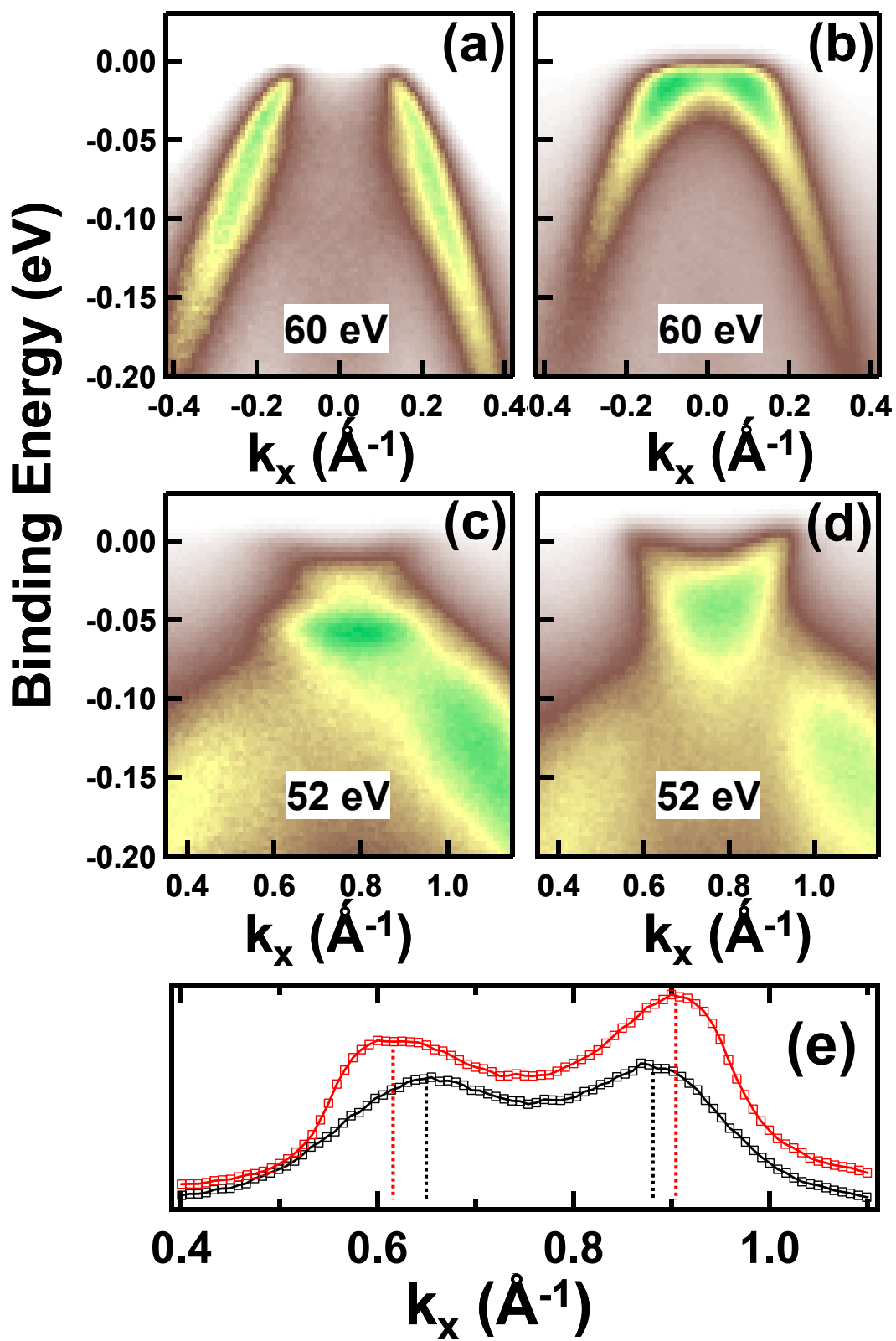

Fig. 1.  (Color Online) Image plots of data collected with in-plane photon polarization along the ΓX direction (i.e. [100]) at the Γ point in the parent $BaFe_2As_2$ (a) and doped $BaFe_{1.8}Co_{0.2}As_2$ (b) compounds, and at the M point for the parent (c) and doped compound (d), respectively. The numbers in the panels denote the photon energy used. Panel (e) shows the MDC curves at zero binding energy for the parent (black line, bottom curve) and doped (red line, top curve) compounds corresponding to the image plots in (c) and (d), respectively. Upon Co substitution, the hole pocket at Γ becomes smaller, while the electron pocket at M becomes larger, indicating that Co substitution induces an upward shift of the chemical potential, as expected for electron doping.

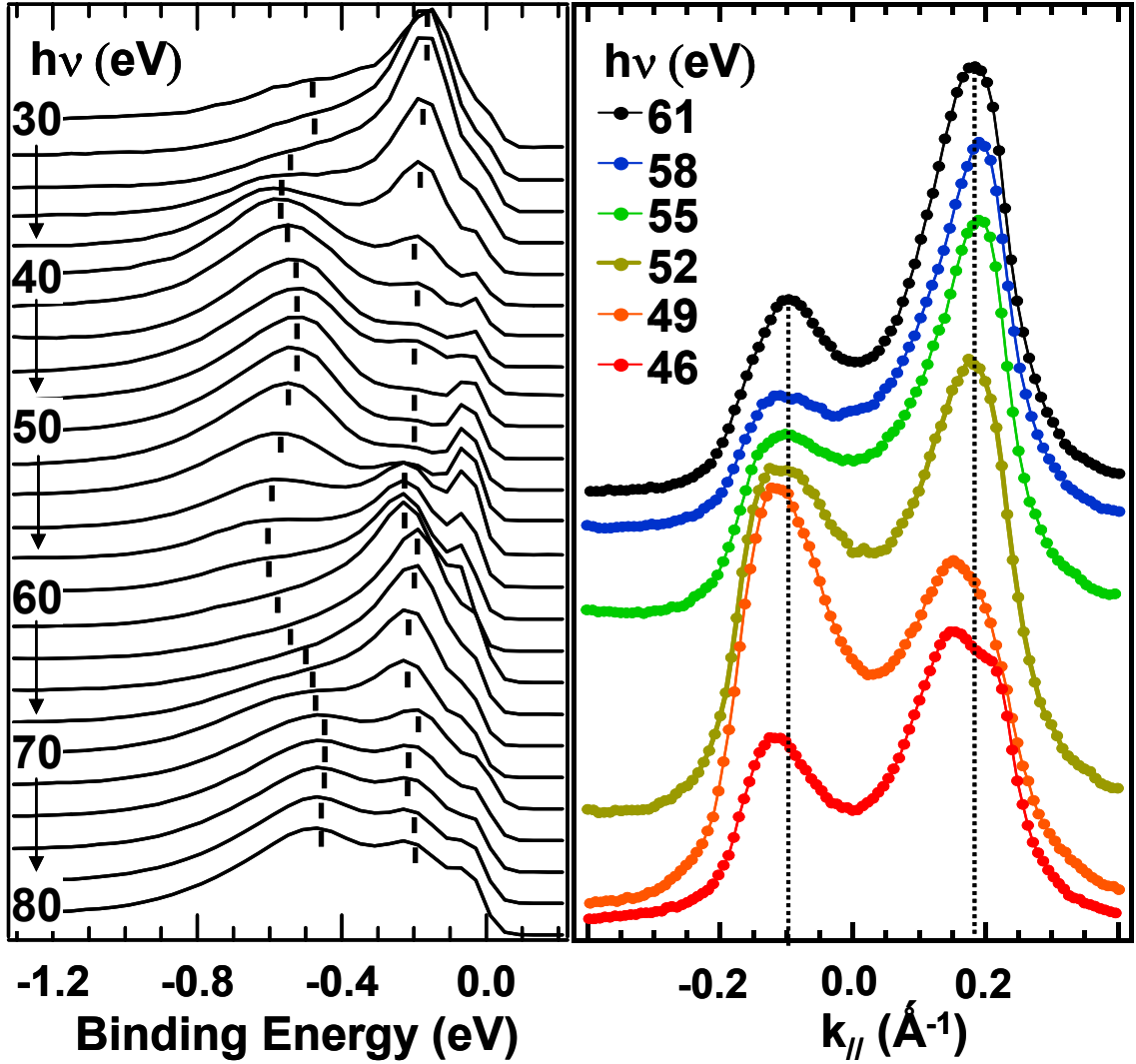

Fig. 2. (Color Online) (a) $BaFe_{1.8}Co_{0.2}As_2$ valence band spectra collected in normal emission with different photon energies. The spectra have been collected in the angular mode with the analyzer slit along the ΓM direction. The photon polarization was mixed, with both in-plane and out-of-plane components. The spectra have been integrated in a window of ± 1° around the Γ point. The dispersion of the structure at ≈ 600 meV is indicative of a three-dimensional character of the band structure. (b) MDC curves at zero binding energy for $BaFe_{1.8}Co_{0.2}As_2$ collected at the M point (electron pocket) with different photon energies (top, hν=61 eV; bottom, hν= 46 eV) along the GX [100] direction and in-plane polarizations. The origin of the horizontal axis has been set to the $k_x$ position of the M point (≈ 0.79 Å$^{-1}$).

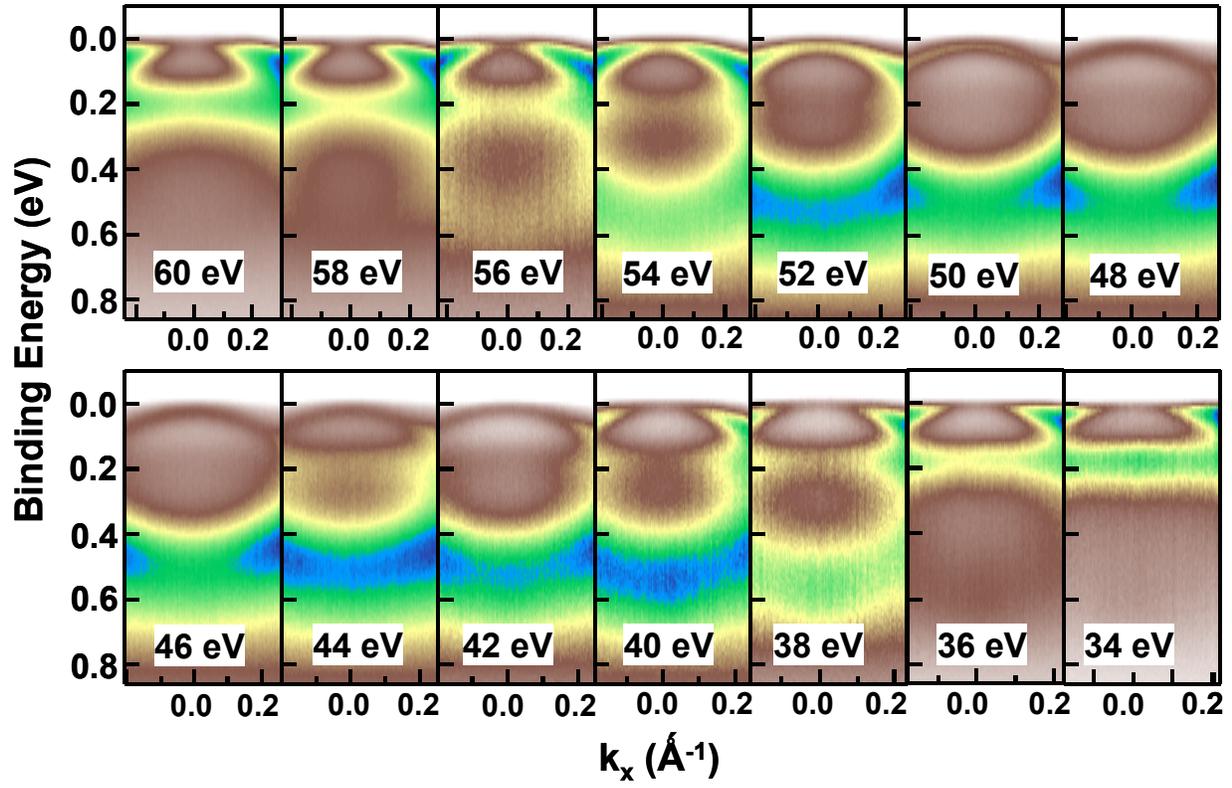

Fig. 3. (Color Online) Photon energy dependence of spectra collected in BaFe$_{1.8}$Co$_{0.2}$As$_2$ at the Γ point along the ΓX direction. The numbers in the insets denote the photon energy used. The photon polarization was mixed, with both in-plane and out-of-plane components. The photon energy dependence of the Fermi crossing points at zero binding energy is evidence of the three-dimensional character of the FS topology.

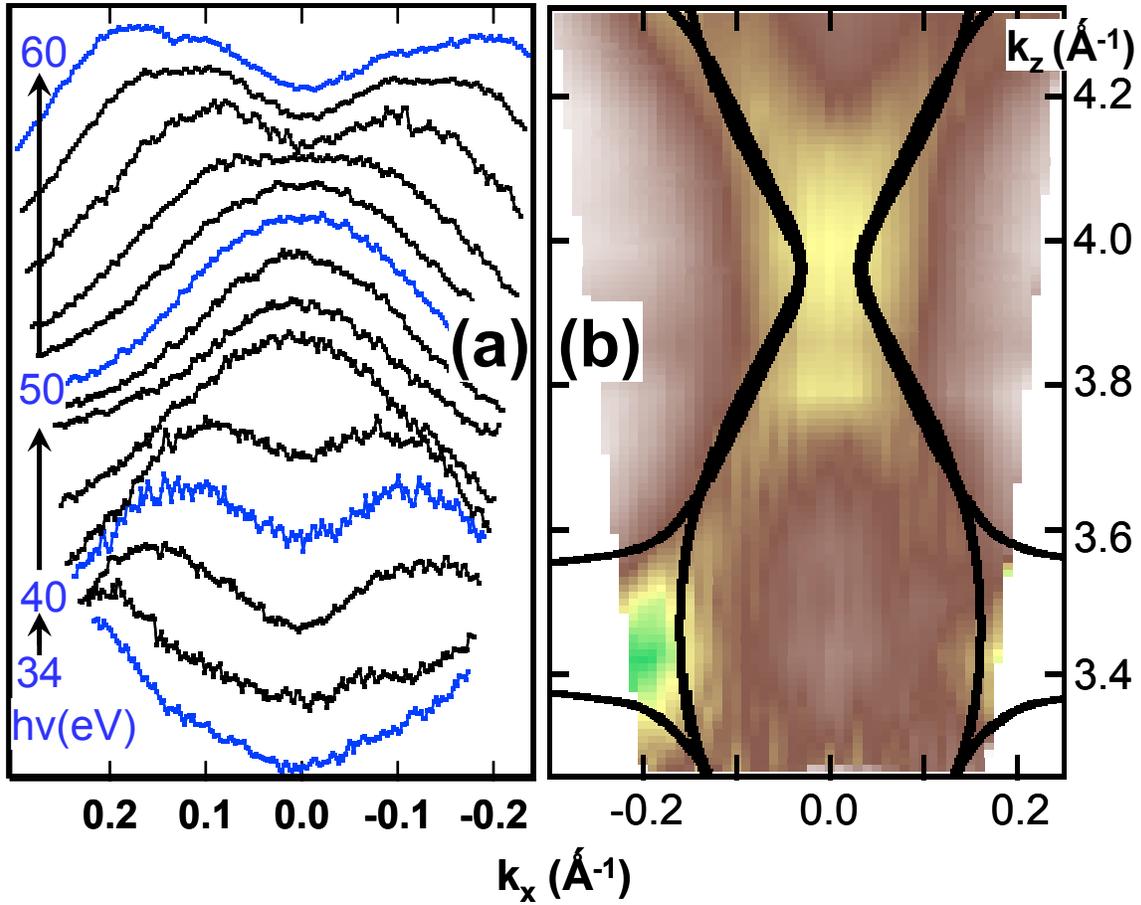

Fig. 4. (Color Online) Analysis of the spectra shown in Fig. 3. (a) MDC curves at $E_F$, and (b) intensity at $E_F$ plotted in color scale as a function of the in plane momentum $k_x$ and vertical momentum $k_z$. The plot illustrates the three-dimensional character of the Fermi surface topology by showing how the FS around the Γ point consists of a cylinder with small cross sectional area that flares out in proximity of the Z point. The Γ and Z points correspond to ≈ 3.88 Å$^{-1}$ and 3.39 Å$^{-1}$ on the vertical $k_z$ scale. The continuous black line in (b) denotes the $k_z$ dispersion calculated for $BaFe_{1.8}Co_{0.2}As_2$ with LDA in the virtual crystal approximation.

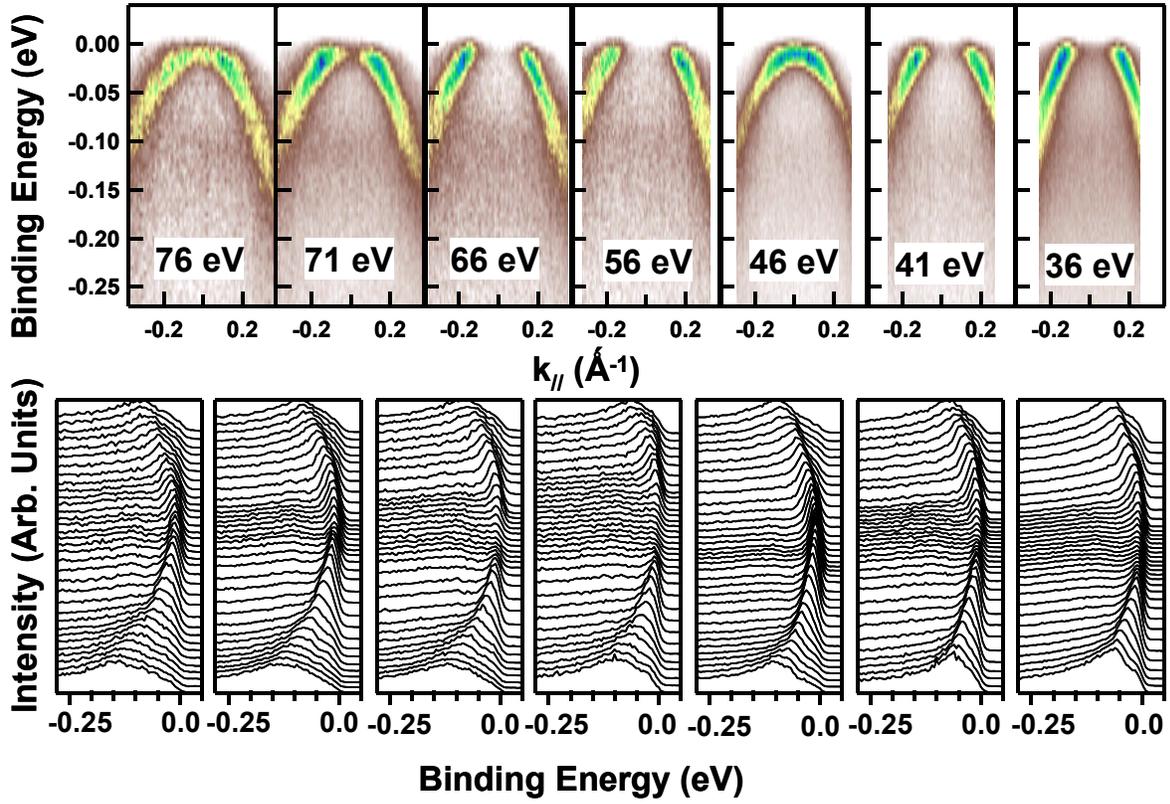

Fig. 5. (Color Online) Photon energy dependence of spectra collected in the parent compound BaFe$_2$As$_2$ at the $\Gamma$ point along the $\Gamma$X direction. The panels in the top and bottom rows denote the image plot and the corresponding energy distribution curves. The numbers in the insets denote the photon energy used. The photon polarization was in-plane. The photon energy dependence of the Fermi crossing points at zero binding energy is evidence of the three-dimensional character of the FS topology.

[22] Photon energies of 50 and 80 eV correspond to $k_z$ values of 3.99 Å$^{-1}$ and 4.88 Å$^{-1}$, respectively, whose difference is close to the expected periodicity along $k_z \approx 4\pi/c = 0.968$ Å$^{-1}$.